\documentstyle[doublespace,prl,epsf,aps,floats]{revtex}  
\input psfig.sty
\def\et{$E_{\rm T}$}                          
\def\met{\mbox{${\hbox{$E$\kern-0.6em\lower-.1ex\hbox{/}}}_T$}} 
\def\mex{\mbox{${\hbox{$E$\kern-0.6em\lower-.1ex\hbox{/}}}_x$}} 
\def\mey{\mbox{${\hbox{$E$\kern-0.6em\lower-.1ex\hbox{/}}}_y$}} 
\def\mexy{\mbox{${\hbox{$E$\kern-0.6em\lower-.1ex\hbox{/}}}_{x,y}$}} 
 
\def\gevcc{GeV/$c^2~$}                   
\def\gevc{GeV/$c$}                       
\def\D0{D\O}                            
\begin{document}
\lefthyphenmin=2
\righthyphenmin=3

%
%

\title{Search for Long-Lived Doubly-Charged Higgs Bosons \\ 
 in $p \bar{p}$ Collisions at $\sqrt{s} = 1.96$ TeV}

\maketitle

\font\eightit=cmti8
\def\r#1{\ignorespaces $^{#1}$}
\hfilneg
\begin{sloppypar}
\noindent 
D.~Acosta,\r {16} J.~Adelman,\r {12} T.~Affolder,\r 9 T.~Akimoto,\r {54}
M.G.~Albrow,\r {15} D.~Ambrose,\r {15} S.~Amerio,\r {42}
D.~Amidei,\r {33} A.~Anastassov,\r {50} K.~Anikeev,\r {15}
% ...
\end{sloppypar}
\vskip .026in
\begin{center}
(CDF Collaboration)
\end{center}

\vskip .026in
\begin{center}
\r 1  {\eightit Institute of Physics, Academia Sinica, Taipei, Taiwan 11529,
Republic of China} \\
\r 2  {\eightit Argonne National Laboratory, Argonne, Illinois 60439} \\
\r 3  {\eightit Institut de Fisica d'Altes Energies, Universitat Autonoma
de Barcelona, E-08193, Bellaterra (Barcelona), Spain} \\
\r 4  {\eightit Istituto Nazionale di Fisica Nucleare, University of Bologna,
I-40127 Bologna, Italy} \\
\r 5  {\eightit Brandeis University, Waltham, Massachusetts 02254} \\
\r 6  {\eightit University of California, Davis, Davis, California  95616} \\
\r 7  {\eightit University of California, Los Angeles, Los
Angeles, California  90024} \\
\r 8  {\eightit University of California, San Diego, La Jolla, California  92093} \\
\r 9  {\eightit University of California, Santa Barbara, Santa Barbara, California
93106} \\
\r {10} {\eightit Instituto de Fisica de Cantabria, CSIC-University of Cantabria,
39005 Santander, Spain} \\
\r {11} {\eightit Carnegie Mellon University, Pittsburgh, PA  15213} \\
\r {12} {\eightit Enrico Fermi Institute, University of Chicago, Chicago,
Illinois 60637} \\
\r {13}  {\eightit Joint Institute for Nuclear Research, RU-141980 Dubna, Russia}
\\
\r {14} {\eightit Duke University, Durham, North Carolina  27708} \\
\r {15} {\eightit Fermi National Accelerator Laboratory, Batavia, Illinois
60510} \\
\r {16} {\eightit University of Florida, Gainesville, Florida  32611} \\
\r {17} {\eightit Laboratori Nazionali di Frascati, Istituto Nazionale di Fisica
               Nucleare, I-00044 Frascati, Italy} \\
\r {18} {\eightit University of Geneva, CH-1211 Geneva 4, Switzerland} \\
\r {19} {\eightit Glasgow University, Glasgow G12 8QQ, United Kingdom}\\
\r {20} {\eightit Harvard University, Cambridge, Massachusetts 02138} \\
\r {21} {\eightit Division of High Energy Physics, Department of
Physics, University of Helsinki and Helsinki Institute of Physics,
FIN-00044, Helsinki, Finland}\\
\r {22} {\eightit Hiroshima University, Higashi-Hiroshima 724, Japan} \\
\r {23} {\eightit University of Illinois, Urbana, Illinois 61801} \\
\r {24} {\eightit The Johns Hopkins University, Baltimore, Maryland 21218} \\
\r {25} {\eightit Institut f\"{u}r Experimentelle Kernphysik,
Universit\"{a}t Karlsruhe, 76128 Karlsruhe, Germany} \\
\r {26} {\eightit High Energy Accelerator Research Organization (KEK), Tsukuba,
Ibaraki 305, Japan} \\
\r {27} {\eightit Center for High Energy Physics: Kyungpook National
University, Taegu 702-701; Seoul National University, Seoul 151-742; and
SungKyunKwan University, Suwon 440-746; Korea} \\
\r {28} {\eightit Ernest Orlando Lawrence Berkeley National Laboratory,
Berkeley, California 94720} \\
\r {29} {\eightit University of Liverpool, Liverpool L69 7ZE, United Kingdom} \\
\r {30} {\eightit University College London, London WC1E 6BT, United Kingdom} \\
\r {31} {\eightit Massachusetts Institute of Technology, Cambridge,
Massachusetts  02139} \\
\r {32} {\eightit Institute of Particle Physics: McGill University,
Montr\'{e}al, Canada H3A~2T8; and University of Toronto, Toronto, Canada
M5S~1A7} \\
\r {33} {\eightit University of Michigan, Ann Arbor, Michigan 48109} \\
\r {34} {\eightit Michigan State University, East Lansing, Michigan  48824} \\
\r {35} {\eightit Institution for Theoretical and Experimental Physics, ITEP,
Moscow 117259, Russia} \\
\r {36} {\eightit University of New Mexico, Albuquerque, New Mexico 87131} \\
\r {37} {\eightit Northwestern University, Evanston, Illinois  60208} \\
\r {38} {\eightit The Ohio State University, Columbus, Ohio  43210} \\
\r {39} {\eightit Okayama University, Okayama 700-8530, Japan}\\
\r {40} {\eightit Osaka City University, Osaka 588, Japan} \\
\r {41} {\eightit University of Oxford, Oxford OX1 3RH, United Kingdom} \\
\r {42} {\eightit University of Padova, Istituto Nazionale di Fisica
          Nucleare, Sezione di Padova-Trento, I-35131 Padova, Italy} \\
\r {43} {\eightit University of Pennsylvania, Philadelphia,
        Pennsylvania 19104} \\
\r {44} {\eightit Istituto Nazionale di Fisica Nucleare Pisa, Universities 
of Pisa, Siena and Scuola Normale Superiore, I-56127 Pisa, Italy} \\
\r {45} {\eightit University of Pittsburgh, Pittsburgh, Pennsylvania 15260} \\
\r {46} {\eightit Purdue University, West Lafayette, Indiana 47907} \\
\r {47} {\eightit University of Rochester, Rochester, New York 14627} \\
\r {48} {\eightit The Rockefeller University, New York, New York 10021} \\
\r {49} {\eightit Istituto Nazionale di Fisica Nucleare, Sezione di Roma 1,
University di Roma ``La Sapienza," I-00185 Roma, Italy}\\
\r {50} {\eightit Rutgers University, Piscataway, New Jersey 08855} \\
\r {51} {\eightit Texas A\&M University, College Station, Texas 77843} \\
\r {52} {\eightit Texas Tech University, Lubbock, Texas 79409} \\
\r {53} {\eightit Istituto Nazionale di Fisica Nucleare, University of Trieste/\
Udine, Italy} \\
\r {54} {\eightit University of Tsukuba, Tsukuba, Ibaraki 305, Japan} \\
\r {55} {\eightit Tufts University, Medford, Massachusetts 02155} \\
\r {56} {\eightit Waseda University, Tokyo 169, Japan} \\
\r {57} {\eightit Wayne State University, Detroit, Michigan  48201} \\
\r {58} {\eightit University of Wisconsin, Madison, Wisconsin 53706} \\
\r {59} {\eightit Yale University, New Haven, Connecticut 06520} \\
\end{center}

%
%
\begin{abstract}
 We present a search for long-lived doubly-charged Higgs bosons
  ($H^{\pm \pm}$), with signatures of high ionization energy loss 
 and muon-like penetration. 
 We use 292  pb$^{-1}$ of data collected in 
 {\mbox{$p\bar p$}}\ collisions at {\mbox{$\sqrt{s}$ =\ 1.96\ TeV }} by
 the CDF II detector at the Fermilab Tevatron. 
 Observing no evidence of long-lived doubly-charged particle production, we  
 exclude $H^{\pm \pm}_L$ and $H^{\pm \pm}_R$ 
 bosons with masses below 133~\gevcc\ and 109~\gevcc, 
 respectively. In the degenerate case we exclude 
 $H^{\pm \pm}$ mass below 146~\gevcc.  All limits are quoted at the 95\% confidence level. 
\end{abstract}


\vfill\eject
The electroweak gauge symmetry of the standard model (SM) is broken by
 the hypothetical Higgs mechanism, thereby imparting masses to the $W$
 and $Z$ bosons,
 the mediators of the weak force. A number of 
 models~\cite{smext,lrsym,lighth++,lightstableh++} 
 extend the SM Higgs sector to include additional symmetries. For instance,
 the left-right symmetric model~\cite{lrsym} postulates a right-handed version
 of the weak interaction, whose gauge symmetry is spontaneously broken
 at a high mass scale, leading to the parity-violating SM. This model
  is supported by recent data on neutrino
 oscillations~\cite{oscillations}, and explains small neutrino
  masses~\cite{lightnu}.
 The model generally requires a Higgs triplet 
 containing a doubly-charged Higgs boson ($H^{\pm \pm}$),
 which could be
 light in the minimal supersymmetric left-right model~\cite{lighth++,lightstableh++}.
 Discovery of the $H^{\pm \pm}$ boson
 would not only shed light on the Higgs mechanism, 
 but also provide evidence for new symmetries beyond the SM. Grand unified
  theories containing Higgs triplets and
 their relevance for neutrino masses and mixing are
 reviewed in~\cite{recent-reviews}, while ``Little Higgs'' models that
 ameliorate the heirarchy and fine-tuning problems of the SM are 
 reviewed in~\cite{littleHiggs}. 

 The dominant production mode at
 the Tevatron is $p\bar{p} \rightarrow \gamma^*/Z + X \rightarrow H^{++} 
 H^{--} + X $, whose cross section at tree level is specified by the quantum
 numbers and the mass  $(m_{H^{\pm \pm}})$ of the $H^{\pm \pm}$ boson. The
 partial width in the leptonic decay modes is given by
 $\Gamma_{ll'} = h_{ll'}^2 m_{H^{\pm \pm}} / (8 \pi)$, where $h_{ll'}$ are  
 phenomenological couplings. 
 In a previous Letter~\cite{dchiggsSS}, we published the most stringent
 $H^{\pm \pm}$ mass limits from direct searches 
 in the $ee$, $e \mu$ and $\mu \mu$ decay
 channels for $0.5 > h_{ll'} >10^{-5}$. 
 In this Letter, we discuss the case 
 where the $H^{\pm \pm}$ boson lifetime ($\tau$) is long ($c \tau 
 > 3$m, corresponding to $h_{ll'} < 10^{-8}$), resulting in the $H^{\pm \pm}$ 
 boson decaying outside the CDF detector~\cite{intra}. A supersymmetric
 left-right model~\cite{lightstableh++}
  has predicted a light $H^{\pm \pm}$ boson with $ B-L=0$, 
 where $B$ and $L$ represent baryon number and lepton number respectively, 
 resulting in  $h_{ll'} = 0$ and a long lifetime~\cite{dchiggsSS}. The LEP experiments
have set limits on a long-lived 
 $H^{\pm \pm}$ boson~\cite{opal,delphi}, with the best limit
coming from the DELPHI experiment~\cite{delphi}, excluding
 $m_{H^{\pm \pm}} < 99.6$~\gevcc\ (99.3~\gevcc) at the
  95\% confidence level (C.L.) 
 for $H^{\pm \pm}$ bosons with couplings to
 left- (right-)handed leptons.
 Our search for pair-production of long-lived, doubly-charged particles is
   based on the signatures of increased ionization
 energy loss and muon-like penetration of
  shielding (due to their large mass).  We set
  the most stringent
 $H^{\pm \pm}$ mass
 limits in the context of the left-right symmetric model. 

 This analysis  uses $292 \pm 18$~pb$^{-1}$ 
 of data collected by the CDF II detector~\cite{tdr}
  in $p \bar{p} $ collisions at $\sqrt{s} = 1.96$~TeV at the 
 Tevatron. The detector consists of a cylindrical magnetic spectrometer with
 silicon and drift chamber trackers,
 surrounded by a time-of-flight system,
 pre-shower detectors, electromagnetic (EM) and hadronic calorimeters,
 and muon detectors. The 
 central drift chamber
  (COT)~\cite{cot},  central 
 calorimeter~\cite{run1detector} and the muon detectors~\cite{muondetectors},
 covering the region
 $| \eta | < 1$~\cite{coords}, are used in this analysis.
  The COT and calorimeter provide 
 ionization information in addition  to tracking and identification of
 penetrating particles. 
 
 We use an inclusive muon trigger requiring a COT track with 
 transverse momentum
  $p_{\rm T} > 18$~\gevc\ \cite{coords}, and a matching track segment
 in the central
 muon chambers. In the offline analysis, we search for $H^{++}H^{--}$  
 pair-production by requiring two 
 COT tracks, each with $p_{\rm T} > 20$~\gevc, 
 beam impact parameter~$<$~2~mm and at least 30 (out of a maximum of 96)
 sense wire hits. 
 At least one of the tracks  is required to have a matching muon chamber
 segment.  
  We also 
 require their isolation $I_{0.4} < 0.1$, where $I_{0.4}$ is the ratio
 of the total
 calorimeter \et\ \cite{coords} around the track within a cone of radius 
 $R \equiv
 \sqrt { (\Delta \eta) ^ 2 + (\Delta \phi) ^ 2} = 0.4$ to the track 
 $p_{\rm T}$~\cite{coords}. Energy deposited by the particle is excluded
 from the calculation of $I_{0.4}$. Finally, we tag and reject
 cosmic ray tracks using an algorithm based on COT
  hit-timing~\cite{cotcosmic}, whose efficiency is measured to be
 $100^{+0.0}_{-0.8}$\%
 for collider muons and leaves negligible cosmic ray
 contamination. 
 
We use $Z \rightarrow \mu \mu$
 events that were triggered by one of the muons
      to measure trigger and offline identification efficiencies of the other
      muon. The track selection efficiency
 is $(93.6 \pm 0.2)$\%, and the 
 efficiency for one of the two $H^{\pm \pm}$ bosons
  to satisfy the 
 muon trigger and matching-segment requirements  
 is  $(96.8 \pm 0.7)$\%. The effect of 
 increased multiple-scattering of doubly-charged particles  
 is investigated  by comparing  the segment matching 
 efficiency for muons from $Z$ boson decays with that for
  lower-$p_{\rm T}$ muons from  
$\Upsilon$ decays. The small ($\approx$0.5\%) difference, when scaled 
 as $p_{\rm T}^{-1}$ to the large $p_{\rm T}$ of 
 $H^{\pm \pm}$ tracks, predicts
 a negligible ($\approx$0.2\%)
  correction. About 3\% of
   $H^{\pm \pm}$ particles are expected to be sufficiently slow ($\beta < 0.4$)
 to have a reduced  efficiency due to delayed hits, for a net
 efficiency loss of 0.4\%. 
  A correction is applied  
 to the track selection efficiency for $H^{\pm \pm}$ bosons passing
 near a calorimeter tower edge and depositing 
 a large ionization energy signal in an adjacent tower. This effect, caused
 by the resolution of the track extrapolation, leads to the $H^{\pm \pm}$ boson
 candidate failing
 the isolation requirement. This geometrical correction results in an 
 overall $H^{\pm \pm}$ track selection
 efficiency of $(89 \pm 4)$\%. 

 The charge collected by each COT wire is proportional
 to the ionization deposited by the particle per unit length (d$E$/d$x$), and
  is encoded in the width of the digital pulse
 generated by the front-end electronics~\cite{cot}. 
 Offline corrections are applied for the 
 electronics response, track polar angle, COT high voltage, drift distance,
 drift direction with respect to track direction, gas  
 pressure, attenuation along the sense wire, radial 
 location of the sense wire, 
  and time.
  The mean number of hits
 on our selected tracks is 85. The mean $(w)$
 of the
 lower 80\% of the corrected 
 widths of all recorded hits of a track is used as a measure 
 of its ionization energy loss. The use of the truncated mean reduces 
 the  sensitivity to 
 Landau fluctuations. 

\begin{figure}[!htbp]
\begin{center}
\epsfysize = 5.2cm
\epsffile{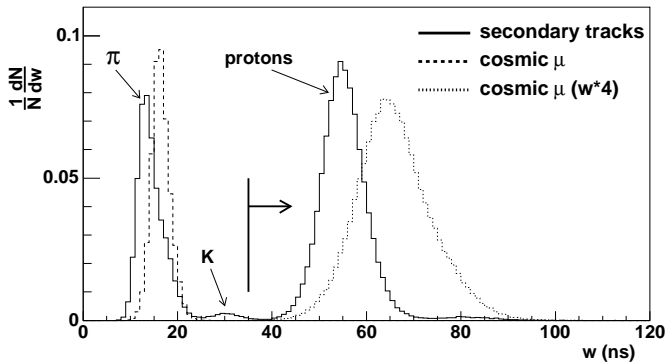}
\vspace*{1mm}
\caption{The distribution of the COT d$E$/d$x$ variable $w$ for positively-charged
 secondary~\protect\cite{secondary} 
tracks in the momentum range
 of $300-350$~MeV/$c$ (solid), for high$-p_{\rm T}$ cosmic ray muons (dashed), and the 
 expectation for $H^{\pm \pm}$
 tracks (dotted). The latter is modelled by quadrupling the $w$ measurements of cosmic ray muons. The arrow indicates the signal selection region.}
\label{cotdedx}
\end{center}
\end{figure}

 The most probable d$E$/d$x$ for a 
 minimum-ionizing particle corresponds to $w \approx 15$~ns, as seen from
 the cosmic-ray muon distribution in Fig.~\ref{cotdedx}. For the
 $H^{\pm \pm}$ search we require $w > 35$~ns. The $w$ distribution of 
 the latter
 is modelled by quadrupling the $w$ measurements of cosmic ray muons, as 
 given by
 the (charge)$^2$-dependence of ionization energy loss in the Bethe-Bloch 
 equation. 
We use low-momentum protons from
secondary interactions to measure the efficiency of the d$E$/d$x$ cut on
$H^{\pm \pm}$ tracks,
 which are expected to have similar or greater d$E$/d$x$ than said
protons (see Fig.~\ref{cotdedx}). 
 We obtain a control sample enriched in
 highly-ionizing protons by selecting  low-momentum
 positively-charged secondary~\cite{secondary}
 tracks.  The pion contribution is
 statistically removed by
  subtracting the $w$ distribution of negatively-charged
  secondary tracks. Using the resulting $w$ distribution of protons,
 we measure the $w$ selection 
 efficiency to be $>99.5$\%. 

 We perform two simultaneous searches with ``loose'' and ``tight'' selections
 for highly-ionizing particles. The loose selection, based on the COT
 d$E$/d$x$ measurement
 only, yields the maximum acceptance, while the tight selection
 also requires large EM and hadronic calorimeter signals for confirmation
 of a potential signal. We make the {\it a priori} decision to use the results 
 from the ``loose'' 
 search to quote an upper limit on the signal cross section, and the ``tight''
 search  results to quote a statistically significant observation of signal. 
 The most probable ionization energy signal deposited by muons
 in the EM and hadronic
 calorimeters (referred to as $E_{\rm EM}$ and $E_{\rm had}$ respectively) 
 is 0.3 GeV and 1.7 GeV respectively, for normal incidence. 
 For the tight $H^{\pm \pm}$ search we require $E_{\rm EM} > 0.6$~GeV
  and $E_{\rm had} > 4$~GeV. The efficiency of the calorimeter
 ionization requirements is $(81.1 \pm 0.1)$\%,  measured 
 by quadrupling $E_{\rm EM}$ and $E_{\rm had}$ of a pure cosmic
 ray sample to  model the $H^{\pm \pm}$ energy deposition.  
  
 We calculate the geometric and kinematic acceptance for 
 a pair of $H^{\pm \pm}$ bosons
 using the {\sc PYTHIA}\cite{pythia} generator 
 and a {\sc GEANT}\cite{geant}-based
  detector simulation. The acceptance increases
 from 38.4\% at $m_{H^{\pm \pm}} = 90$~\gevcc\ to 46.8\% 
 at $m_{H^{\pm \pm}} = 160$~\gevcc,
 with the dominant 
 relative systematic uncertainty of 1\% due to parton distribution
  functions (PDFs)~\cite{cteq6}. Systematic uncertainties due to momentum scale and
 resolution are negligible. 
 
\begin{table}
\caption{ Summary of fake rate measurements. The $e$, $\mu$ and $\tau$ fake
 rates and the ``muon fake rates'' for jets
 are quoted as upper limits at the 68\% C.L., since
 no events in the respective control samples 
 pass the $H^{\pm \pm}$ selection cuts.}
\begin{tabular}{c|cc|cc}
\multicolumn{1}{c}{source} & \multicolumn{2}{c|}{loose search}&\multicolumn{2}{c}{tight search}\\
 & ``track'' & ``muon'' & ``track'' &  ``muon'' \\
\hline
jet $(\times 10^{-4})$ & $3.2^{+5.0}_{-2.9} $  & $ < 0.05 $  & $0.28^{+0.04}_{-0.05} $ & 
$ < 0.05 $ \\
$e$ $(\times 10^{-6})$ & $<4$ & $< 0.00009 $    & $<0.05$  &  $ < 0.00002  $ \\
$\mu$ $(\times 10^{-6})$ & $< 7  $  & $< 7 $    &  $< 0.02 $ & $< 0.02  $\\
$\tau$ $(\times 10^{-5})$ & $< 2 $ &  $< 0.002 $    &  $< 2 $   & $< 0.002  $   \\
\end{tabular}
\label{fakerates}
\end{table}

 Backgrounds arise
 from (1) jets fragmenting into high-$p_{\rm T}$ tracks,
 (2) $Z \rightarrow ee$, (3) $Z \rightarrow \mu \mu$, and (4)
 $Z \rightarrow \tau \tau$ where at least one $\tau$ decays hadronically.
 The backgrounds are a result of muon misidentification and 
 d$E$/d$x$ mismeasurement, which can arise from overlapping particles. 
 Each background is estimated by multiplying the number of 
 misidentifiable events by the product of the appropriate misidentification
 probabilities (fake rates). Fake rates are measured with
  and without 
 the requirement of a matching muon chamber segment. 
We refer to these as the ``muon fake rate"
   and ``track fake rate", respectively. 
A fake rate is defined as
 the probability that a track (or muon) 
passing certain loose identification cuts also 
 satisfies the analysis cuts. For jets, electrons and $\tau$'s, the muon 
 fake rate is
 obtained by multiplying the track fake rate by the estimated probability of 
 mis-matching 
 a muon chamber segment to the track.

The track fake rate and muon fake rate for jets are measured from
 jet-triggered data and muon-triggered data, respectively. 
 The variation of the fake rates with $p_{\rm T}$ and jet proximity is 
 taken as a measure of systematic uncertainty.  The number of
 misidentifiable jet events is given by the number of
 muon-triggered data events
 containing a loosely-selected muon and another loosely-selected track.
 Fake rates
 for electrons and hadronically decaying $\tau$'s are estimated from 
 the {\sc GEANT}-based detector simulation. These fake
 rate measurements are limited by Monte Carlo statistics, as no Monte Carlo
 events pass the $H^{\pm \pm}$ selection cuts. The number of misidentifiable $Z \rightarrow ee$ events
 is obtained from the $Z \rightarrow ee$ data sample, corrected for electron efficiencies and
 normalized to the luminosity of the 
 muon-triggered signal sample. The number of $Z \rightarrow \tau \tau$ misidentifiable events
 is obtained from the number of $Z \rightarrow \mu\mu$ events observed in the data, assuming $\mu - \tau$ 
 universality, and correcting for muon efficiencies. Finally, fake rates for muons
 are measured from a pure sample of cosmic rays, which are again statistically
 limited as no events pass the $H^{\pm \pm}$ selection cuts. The number of misidentifiable events
 is given by the number of $Z \rightarrow \mu\mu$ data events selected with the loose cuts.  
 Table~\ref{fakerates} summarizes the fake rate measurements, and
 Table~\ref{backgrounds} summarizes the resulting background estimates. 
\begin{table}
\caption{ Summary of the estimated number of background events (quoted as 68\% C.L. upper limits)
  and the observed number
 of events in the data.  }
\begin{tabular}{c|c|c}
background & loose search &  tight search \\
\hline
jet   &  $< 3   \times 10^{-5}$   & $< 3  \times 10^{-6}$  \\
$Z \rightarrow ee$ &    $< 1 \times 10^{-11}$   &  $< 2 \times 10^{-14}$ \\
$Z \rightarrow \mu \mu$ &  $< 4 \times 10^{-7}$  &  $< 4 \times 10^{-12}$  \\
$Z \rightarrow \tau \tau$ &  $< 8 \times 10^{-9}$ &  $< 8 \times 10^{-9}$ \\
\hline
\hline
data & 0 & 0 \\
\end{tabular}
\label{backgrounds}
\end{table}
 
\begin{figure}[!htbp]
\begin{center}
\epsfysize = 5.2cm
\epsffile{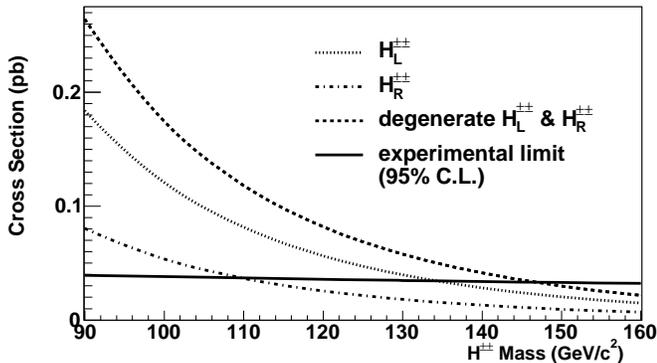}
\vspace*{1mm}
\caption{The comparison of the experimental cross section upper limit with the 
 theoretical next-to-leading order cross section~\protect\cite{spira}
 for pair production of 
 $H^{\pm \pm}$ bosons. The theoretical cross sections are computed separately
 for bosons with left-handed ($H^{\pm \pm}_L$) and right-handed ($H^{\pm \pm}_R$)
 couplings, and summed for the case that their masses are degenerate.}
\label{crossSectionLimit}
\end{center}
\end{figure}

 No $H^{++} H^{--}$ 
 candidate events are found in the data. The null result is used to
  set upper limits on the number of signal events (3.2 at the 95\% C.L.)
  and the $H^{\pm \pm}$ pair 
 production cross section using a Bayesian~\cite{bayes}
  approach, with a flat prior for the signal cross section
 and Gaussian priors for the uncertainties on 
 acceptance, background and integrated luminosity~(6\%)~\cite{luminosity}.
 The 95\% C.L. upper limit on the cross section
 (which varies from 39.7~fb at $m_{H^{\pm \pm}} = 90$~\gevcc\ to 32.6~fb at
 $m_{H^{\pm \pm}} = 160$~\gevcc, see Fig.~\ref{crossSectionLimit})
 is converted into an $H^{\pm \pm}$ 
 mass limit by comparing to the theoretical $p \bar{p} \rightarrow \gamma^* / Z
 + X  \rightarrow H^{++} H^{--} + X$ cross section at 
 next-to-leading order~\cite{spira} using the CTEQ6~\cite{cteq6} set of PDFs. 
 We include 
 uncertainties in the theoretical cross
 sections due to PDFs (5\%)~\cite{cteq6} and higher-order QCD corrections 
 (7.5\%)~\cite{spira} in the 
 extraction of the mass limit. The theoretical cross sections are computed 
 separately for $H^{\pm \pm}_L$ and $H^{\pm \pm}_R$ bosons that couple to
 left- and right-handed particles respectively. When only
 one of these states is accessible, we exclude the long-lived 
 $H^{\pm \pm}_L$ boson below a mass of 133~\gevcc\ and the long-lived
 $H^{\pm \pm}_R$ boson below a mass of 109~\gevcc, both at the 95~\% C.L.
 When the two states are degenerate in mass, we exclude 
  $m_{H^{\pm \pm}} < 146$~\gevcc\ at the 95~\% C.L.

In conclusion, we have searched for long-lived doubly-charged particles using their
 signatures of high ionization and muon-like penetration. No evidence is found
 for
 pair-production of such particles, and we set the individual lower
  limits of 133~\gevcc\ and 109~\gevcc, respectively, on the 
 masses of long-lived $H^{\pm \pm}_L$ and $H^{\pm \pm}_R$ bosons. The mass limit for
 the case of degenerate $H^{\pm \pm}_L$ and $H^{\pm \pm}_R$ bosons is 
146~\gevcc.

We thank M. M$\rm\ddot{u}$hlleitner and M. Spira for calculating the next-to-leading
 order  $H^{\pm \pm}$ production cross section.
We thank the Fermilab staff and the technical staffs of the participating 
institutions for their vital contributions.  This work was supported by the 
U.S. Department of Energy and National Science Foundation; the Italian 
Istituto Nazionale di Fisica Nucleare; the Ministry of Education, Culture, 
Sports, Science and Technology of Japan; the Natural Sciences and Engineering 
Research Council of Canada; the National Science Council of the Republic of 
China; the Swiss National Science Foundation; the A.P. Sloan Foundation; the 
Bundesministerium fuer Bildung und Forschung, Germany; the Korean Science and 
Engineering Foundation and the Korean Research Foundation; the Particle 
 Physics 
and Astronomy Research Council and the Royal Society, UK; the Russian 
Foundation for Basic Research; the Comision Interministerial de Ciencia y 
Tecnologia, Spain; and in part by the European Community's Human Potential 
Programme under contract HPRN-CT-2002-00292, Probe for New Physics.


\begin{references}
\bibitem{coords}
CDF uses a cylindrical coordinate system in which $\phi$ is the azimuthal angle, $r$ is
the radius from the nominal beamline, and $+z$ points in the direction of the proton
beam and is zero at the center of the detector.  The pseudorapidity
 $\eta = -$ln[$\tan(\theta/2)$], where $\theta$ is the polar angle with respect
 to the $z$ axis.  Calorimeter energy (track momentum) measured transverse to
  the beam is denoted
as $E_{\rm T}$ ($p_{\rm T}$). Track $p_{\rm T}$ 
is evaluated from its curvature assuming a 
singly-charged particle, and is half of the true $p_{\rm T}$ for a doubly-charged particle.
\bibitem{secondary}
``Secondary'' tracks are those 
 produced in secondary interactions with detector material, and are selected by requiring
 beam impact parameter
 $> 1$ cm.
\end{references}
\end{document}